\documentclass[twocolumn,english,prb,preprintnumbers,amsmath,amssymb,superscriptaddress]{revtex4}
\usepackage[latin9]{inputenc}
\usepackage{amsmath}
\usepackage{graphicx}

\makeatletter
\@ifundefined{textcolor}{}
{%
 \definecolor{BLACK}{gray}{0}
 \definecolor{WHITE}{gray}{1}
 \definecolor{RED}{rgb}{1,0,0}
 \definecolor{GREEN}{rgb}{0,1,0}
 \definecolor{BLUE}{rgb}{0,0,1}
 \definecolor{CYAN}{cmyk}{1,0,0,0}
 \definecolor{MAGENTA}{cmyk}{0,1,0,0}
 \definecolor{YELLOW}{cmyk}{0,0,1,0}
}

\usepackage{dcolumn}
\usepackage{bm}
\usepackage{color}
\usepackage[final]{pdfpages}

\makeatother

\usepackage{babel}

\begin{document}

\preprint{preprint(\today)}

\title{Hidden magnetism uncovered in charge ordered bilayer kagome material ScV$_{6}$Sn$_{6}$}

\author{Z.~Guguchia}
\email{zurab.guguchia@psi.ch}
\affiliation{Laboratory for Muon Spin Spectroscopy, Paul Scherrer Institute, CH-5232 Villigen PSI, Switzerland}

\author{D.J.~Gawryluk}
\email{dariusz.gawryluk@psi.ch} 
\affiliation{Laboratory for Multiscale Materials Experiments, Paul Scherrer Institut, 5232, Villigen PSI, Switzerland}

\author{Soohyeon~Shin}
\thanks{These authors contributed equally to the paper.}
\affiliation{Laboratory for Multiscale Materials Experiments, Paul Scherrer Institut, 5232, Villigen PSI, Switzerland}

\author{Z.~Hao}
\thanks{These authors contributed equally to the paper.}
\affiliation{Department of Physics, Southern University of Science and Technology, Shenzhen, Guangdong, 518055, China}

\author{C.~Mielke III}
\affiliation{Laboratory for Muon Spin Spectroscopy, Paul Scherrer Institute, CH-5232 Villigen PSI, Switzerland}
\affiliation{Physik-Institut, Universit\"{a}t Z\"{u}rich, Winterthurerstrasse 190, CH-8057 Z\"{u}rich, Switzerland}

\author{D.~Das}
\affiliation{Laboratory for Muon Spin Spectroscopy, Paul Scherrer Institute, CH-5232 Villigen PSI, Switzerland}

\author{I.~Plokhikh}
\affiliation{Laboratory for Multiscale Materials Experiments, Paul Scherrer Institut, 5232, Villigen PSI, Switzerland}

\author{L.~Liborio}
\affiliation{Scientific Computing Department, Science \& Technology Facilities Council, Rutherford Appleton Laboratory, Didcot OX11 0QX, United Kingdom}

\author{K.~Shenton}
\affiliation{Scientific Computing Department, Science \& Technology Facilities Council, Rutherford Appleton Laboratory, Didcot OX11 0QX, United Kingdom}

\author{Y.~Hu}
\affiliation{Photon Science Division, Paul Scherrer Institut, CH-5232 Villigen PSI, Switzerland}

\author{V.~Sazgari}
\affiliation{Laboratory for Muon Spin Spectroscopy, Paul Scherrer Institute, CH-5232 Villigen PSI, Switzerland}

\author{M.~Medarde}
\affiliation{Laboratory for Multiscale Materials Experiments, Paul Scherrer Institut, CH-5232 Villigen PSI, Switzerland}

\author{H.~Deng}
\affiliation{Department of Physics, Southern University of Science and Technology, Shenzhen, Guangdong, 518055, China}

\author{Y.~Cai}
\affiliation{Shenzhen Institute for Quantum Science and Engineering, Southern University of Science and Technology, Shenzhen 518055, China}

\author{C.~Chen}
\affiliation{Shenzhen Institute for Quantum Science and Engineering, Southern University of Science and Technology, Shenzhen 518055, China}

\author{Y.~Jiang}
\affiliation{Laboratory for Topological Quantum Matter and Advanced Spectroscopy (B7), Department of Physics,
Princeton University, Princeton, New Jersey 08544, USA}

\author{A.~Amato}
\affiliation{Laboratory for Muon Spin Spectroscopy, Paul Scherrer Institute, CH-5232 Villigen PSI, Switzerland}

\author{M.~Shi}
\affiliation{Photon Science Division, Paul Scherrer Institut, CH-5232 Villigen PSI, Switzerland}

\author{M.Z. Hasan}
\affiliation{Laboratory for Topological Quantum Matter and Advanced Spectroscopy (B7), Department of Physics,
Princeton University, Princeton, New Jersey 08544, USA}
\affiliation{Princeton Institute for the Science and Technology of Materials, Princeton University, Princeton, New Jersey 08540, USA}
\affiliation{Quantum Science Center, Oak Ridge, Tennessee 37831, USA}

\author{J.-X.~Yin}
\affiliation{Department of Physics, Southern University of Science and Technology, Shenzhen, Guangdong, 518055, China}

\author{R.~Khasanov}
\affiliation{Laboratory for Muon Spin Spectroscopy, Paul Scherrer Institute, CH-5232 Villigen PSI, Switzerland}

\author{E.~Pomjakushina}
\affiliation{Laboratory for Multiscale Materials Experiments, Paul Scherrer Institut, 5232, Villigen PSI, Switzerland}

\author{H.~Luetkens}
\affiliation{Laboratory for Muon Spin Spectroscopy, Paul Scherrer Institute, CH-5232 Villigen PSI, Switzerland}

\maketitle

\textbf{Charge ordered kagome lattices have been demonstrated to be intriguing platforms for studying the intertwining of topology, correlation, and magnetism. The recently discovered charge ordered kagome material ScV$_{6}$Sn$_{6}$ does not feature a magnetic groundstate or excitations, thus it is often regarded as a conventional paramagnet. Here, using advanced muon-spin rotation spectroscopy, we uncover an unexpected hidden magnetism of the charge order. We observe a striking enhancement of the internal field width sensed by the muon ensemble, which takes place within the charge ordered state. More remarkably, the muon spin relaxation rate below the charge ordering temperature is substantially enhanced by applying an external magnetic field. Taken together with the hidden magnetism found in $A$V$_{3}$Sb$_{5}$ ($A$ = K, Rb, Cs) and FeGe kagome systems, our results suggest ubiqitous time-reversal symmetry-breaking in charge ordered kagome lattices.}

One of the most scientifically fruitful families of layered systems shown to host intriguing and exotic physics are the kagome lattice materials \cite{Syozi,ZHou,TbNature,GuguchiaCSS,Mazin}, composed of a two-dimensional lattice of corner-sharing triangles. This unique geometry gives rise to frustration, correlated quantum orders, and topology, owing to its special features: the electronic structure hosts a flat band, inflection points called ``van Hove singularities``, and Dirac cones. 

\begin{figure*}[t!]
\centering
\includegraphics[width=0.8\linewidth]{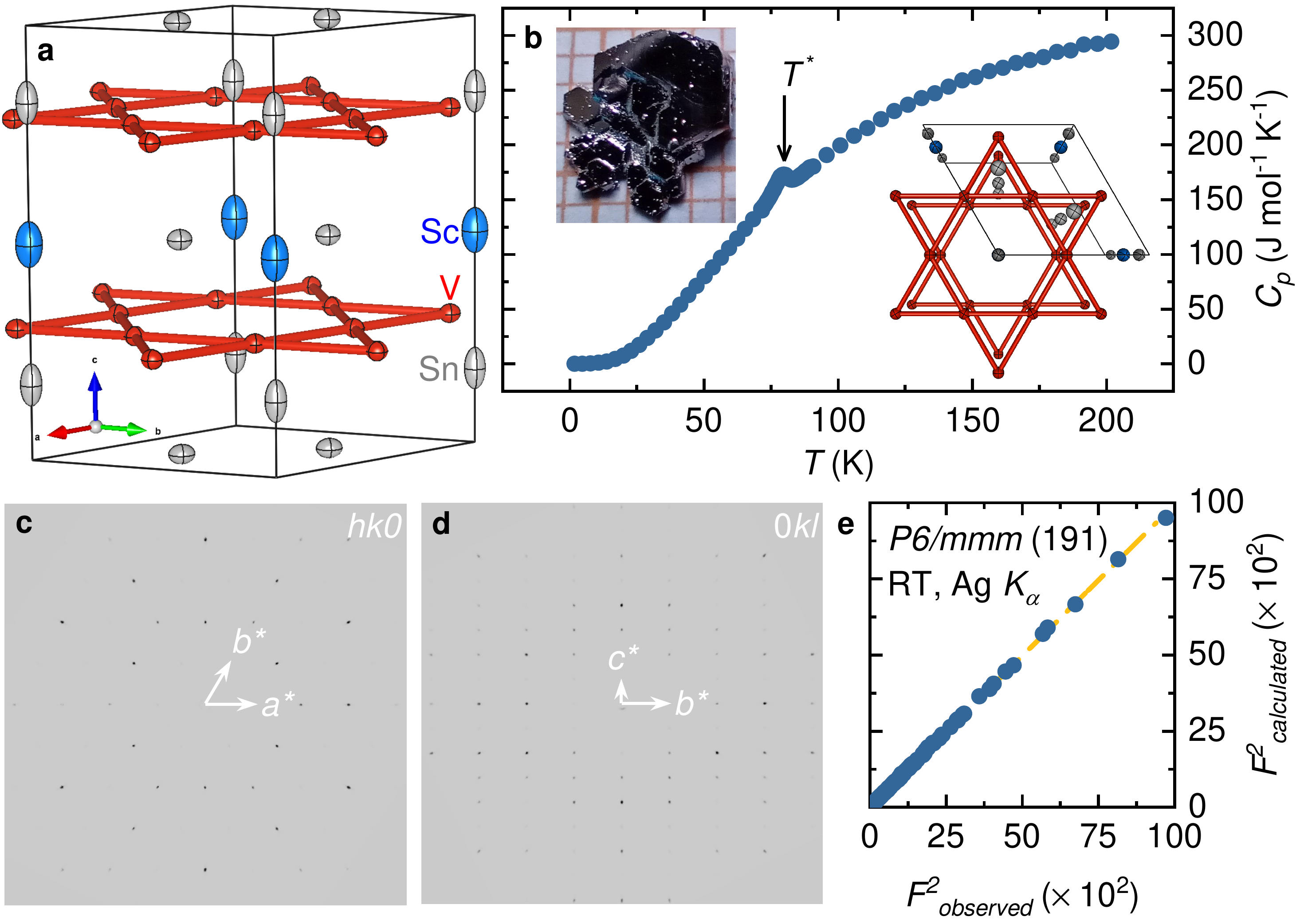}
\vspace{-0.3cm}
\caption{\textbf{Crystal structure of ScV$_{6}$Sn$_{6}$.} 
$\bf{a,}$ Room temperature (RT) crystal structures of ScV$_{6}$Sn$_{6}$ with atomic displacement ellipsoids as determined from laboratory single-crystal X-ray diffraction (Ag $K_{\rm \alpha}$) using the Space Group (SG) $P$6/$mmm$ (191) highlighting the kagome lattice pattern of the vanadium atoms. $\bf{b,}$ Temperature dependence of the heat capacity for the ScV$_{6}$Sn$_{6}$ single crystals.  Insets show a photograph of ScV$_{6}$Sn$_{6}$ single crystal and the crystal structure along c-direction.  $\bf{c,}$ and $\bf{d,}$ Reciprocal $(hk0)$ and $(0kl)$ respectively, lattice planes measured by laboratory X-ray diffraction at RT for a ScV$_{6}$Sn$_{6}$ single crystal, illustrating the agreement between the observed systematic absences and those of the $P$6/$mmm$ SG. $\bf{e,}$ Agreement between the observed and calculated diffraction data at RT, as obtained from structural refinements using the $Olex$$s^{2}$ software \cite{Dolomanov} with $ShelXL$ package \cite{SheldrickL}. Structures were plotted using the $VESTA$ visualization tool \cite{Momma}.}
\label{fig1}
\end{figure*}

The recently discovered $A$V$_{3}$Sb$_{5}$ ($A$ = K, Rb, Cs) \cite{BOrtiz2,BOrtiz3,QYin} family of materials, crystallizing in a structurally perfect two-dimensional kagome net of vanadium atoms, shows the typical kagome band structure and exhibits two correlated orders: high-temperature charge order and a superconducting instability at lower temperatures \cite{YJiang,NShumiya,Wang2021}. The unique feature of $A$V$_{3}$Sb$_{5}$ is the emergence of a time-reversal symmetry breaking chiral charge order with both magnetic \cite{GuguchiaMielke,GuguchiaRVS,KhasanovCVS,LiYu,YXu,YHu,GuoMoll,TNeupert,HasanNature} and electronic anomalies \cite{SYang,FYu}. Theoretically, these features could be explained by a complex order parameter realizing a higher angular momentum state, dubbed unconventional, in analogy to superconducting orders \cite{MDenner,MHChristensen,Balents,Nandkishore,DSong,MortenRafael}. 

$R$V$_{6}$Sn$_{6}$ ($R$ = Sc, Y, or rare earth) \cite{Ganesh,Peng,YHu,Arachchige} is a more recent family of kagome metals that has a similar vanadium structural motif as the $A$V$_{3}$Sb$_{5}$ ($A$ = K, Rb, Cs) compounds. However, ScV$_{6}$Sn$_{6}$ is the only compound among the series of $R$V$_{6}$Sn$_{6}$ that has been found to host charge order below $T^{*}$ ${\simeq}$ 90 K \cite{Arachchige}. Some distinct characteristics \cite{Tianchen,ManuelT,ChengS,YongMing} of charge order such as different propagation vectors and distinct electron-phonon coupling strength \cite{YongMing} have been reported in these two families of kagome metals. However, ScV$_{6}$Sn$_{6}$ shows no superconductivity even after suppression of charge order by hydrostatic pressure \cite{XZhang}. One of the most intruiging questions about the charge order in ScV$_{6}$Sn$_{6}$ is whether or not it breaks time-reversal symmetry.

In this paper, we utilize the combination of zero-field (ZF) and transverse-field (TF) $\mu$SR techniques to probe the $\mu$SR relaxation rates in ScV$_{6}$Sn$_{6}$ as a function of temperature and magnetic field. We detect an enhancement of the muon-spin depolarization rate below the charge order temperature $T^{*}$. The rate is further and significantly enhanced by applying an external magnetic field along the crystallographic $c-$axis, which is indicative of a strong contribution of electronic origin to the muon spin relaxation below the charge ordering temperature. These results point to a complex electronic response strongly intertwined with the charge order in the kagome system ScV$_{6}$Sn$_{6}$ and provide useful insights into the microscopic mechanisms involved in the charge density wave order.

The layered crystal structure of ScV$_{6}$Sn$_{6}$ with its V atoms arranged in a kagome network is shown in Figure 1a.
Figure 1b presents the temperature dependence of the specific heat capacity, $C_{{\rm p}}$. A well pronounced peak in the heat capacity, corresponding to the charge order transition at $T^{*}$ ${\simeq}$ 80 K, is clearly seen, which provides evidence of a bulk phase transition. 
Single crystal X-ray diffraction measurements (see Fig. 1c-e) confirm that the as-grown sample is a high-quality single crystal in the space group $P$6/$mmm$. The lattice parameters $a$ = 5.4739(3) Å and $c$ = 9.1988(7) Å at room temperature are close those reported in previous studies \cite{Arachchige}.



\begin{figure*}[t!]
\centering
\includegraphics[width=1.0\linewidth]{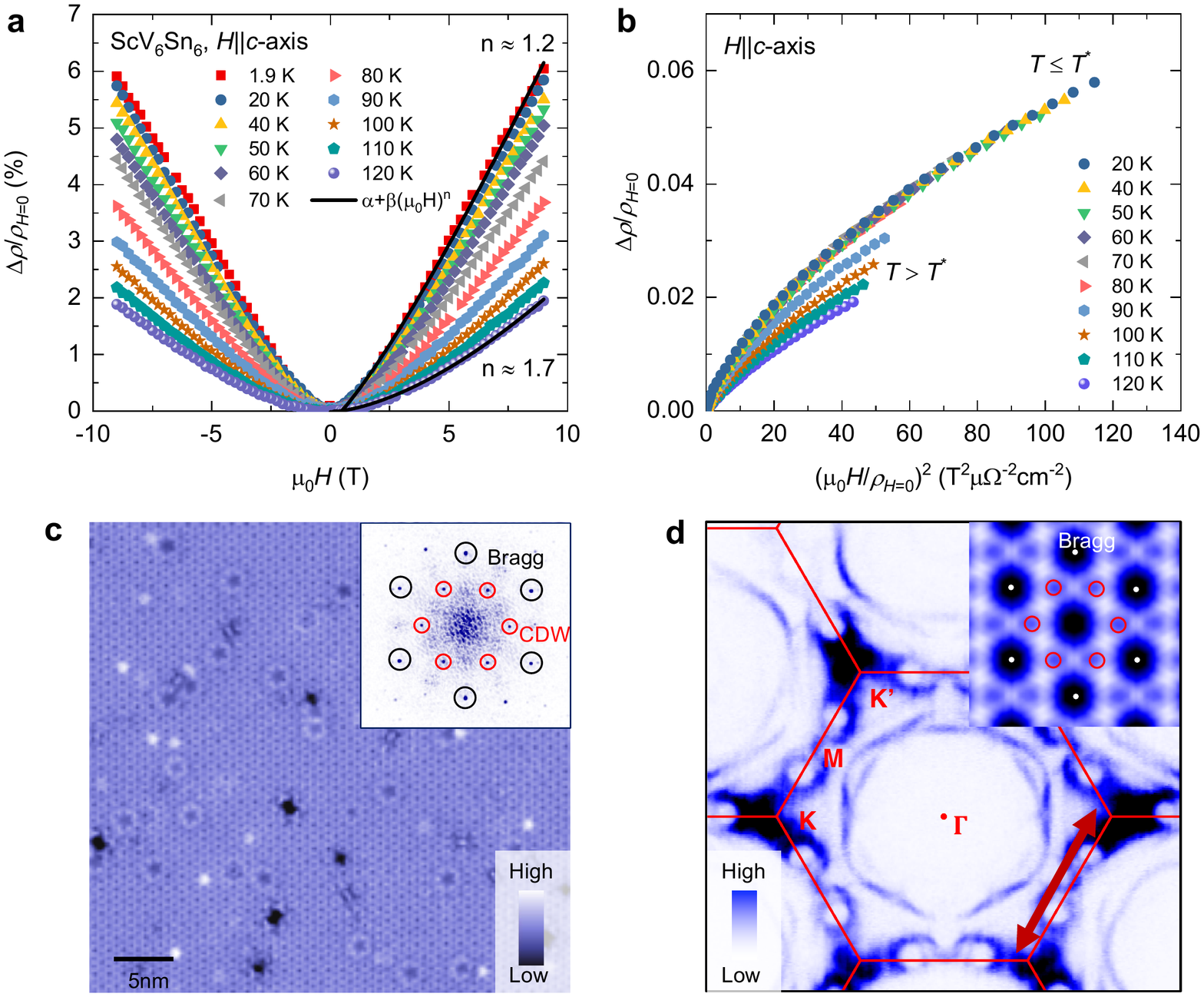}
\vspace{-1.0cm}
\caption{\textbf{Magnetotransport characteristics and charge density wave with an in-plane unit cell enlargement of $\sqrt{3}$ ${\times}$ $\sqrt{3}$ for ScV$_{6}$Sn$_{6}$.} 
$\bf{a,}$ The magnetoresistance measured at various temperatures above and below the charge ordering temperature $T^{*}$ ${\simeq}$ 80 K. Black solid lines represent fits to the data by means of the following equation: ${\Delta}$${\rho}$/${\rho}_{H=0}$ = ${\alpha}$ + ${\beta}$·$H^{n}$  $\bf{b,}$  Kohler plot, ${\Delta}$${\rho}$/${\rho}_{H=0}$ vs (${\mu}_{0}$$H$/${\rho}_{H=0}$)$^{2}$, of the magnetoresistance, plotted from field-sweeps at various temperatures. $\bf{c,}$ STM dI/dV spectroscopic map and corresponding Fourier transform (inset). The red circled marks the $\sqrt{3}$ ${\times}$ $\sqrt{3}$ CDW. Data taken at V = 100 meV, I = 0.5 nA, $T$ = 4.5 K, $V_{\rm mod}$ = 10meV. $\bf{d,}$ Fermi surface obtained by ARPES showing stronger intensities near K points. The inset shows the autocorrelation of the Fermi surface, highlighting the instability with a vector of $\sqrt{3}$ ${\times}$ $\sqrt{3}$ as marked by the red circles. This vector corresponds to the scattering between states at K and $K^{\prime}$ points as marked by the arrow in the main figure. Data taken at 10 K.}
\label{fig1}
\end{figure*}


Magnetotransport data provide further evidence for the charge order transition. Magnetotransport has been shown to be particularly sensitive for detecting the CDW transition, since the magnetoresistance (MR) is a measure of the mean free path integrated over the Fermi surface and can detect a change of the scattering anisotropy and/or a Fermi surface reconstruction reliably. The MR under perpendicular magnetic field in ScV$_{6}$Sn$_{6}$ from 1.9 K to 120 K is shown in Fig. 2a. At high temperatures, MR adopts a standard quadratic dependence with ${\mu}_{0}H$ (${\mu}_{0}H$ is the applied magnetic field) and can be well fitted to a polynomial: ${\Delta}$${\rho}$/${\rho}_{H=0}$ = ${\alpha}$ + ${\beta}$·$H^{n}$ (where ${\alpha}$, ${\beta}$ and $n$ are fitting parameters) with $n$ close to 2. MR shows deviation from quadratic field dependence upon lowering the temperature and becomes nearly linear deep within the charge-ordered state. The exponent is estimated to be $n$ ${\simeq}$ 1.2 at 1.9 K. Linear MR has usually been considered as a hallmark feature of an unconventional quantum state as such behaviour was peviously observed in unconventional superconductors \cite{Giraldo}, topological \cite{Novak} and charge/spin density wave materials \cite{SYang,XWei}. For further insight, we provide the so called Kohler plot, ${\Delta}$${\rho}$/${\rho}_{H=0}$ vs (${\mu}_{0}$$H$/${\rho}_{H=0}$)$^{2}$, which is shown in Fig. 2b. The most striking feature of this plot is that the MR data collapse to one line for all temperatures below $T^{*}$ ${\simeq}$ 80 K while they show a pronounced $T$-dependence above $T^{*}$. This effect may arise from a Fermi surface reconstruction caused by the CDW order.

Our scanning tunneling microscopy (STM) experiments provide direct confirmation for the charge order with an in-plane unit cell enlargement of $\sqrt{3}$ ${\times}$ $\sqrt{3}$ (see Fig. 2c). The angle-resolved photoemission spectroscopy (ARPES) shows that the Fermi surface has stronger intensities at K and $K^{\prime}$ points (see Fig. 2d and also Ref. \cite{YongMing}). The autocorrelation of such Fermi surface data is consistent with instabilities at vectors featuring a $\sqrt{3}$ ${\times}$ $\sqrt{3}$ unit cell enlargement. 

Next, in order to search for any magnetism (static or slowly fluctuating) associated with charge order in ScV$_{6}$Sn$_{6}$, zero-field 
${\mu}$SR experiments have been carried out above and below the charge order temperature $T^{*}$. A schematic overview of the experimental setup with the muon spin forming $45^{\circ}$ with respect to the $c$-axis of the crystal is shown in Figure~3$\bf{a}$. The sample was surrounded by four detectors: Forward (1), Backward (2), Up (3), and Down (4). Figure 3$\bf{b}$ displays the zero-field ${\mu}$SR spectra from detectors 3 \& 4 collected over a wide temperature range. The ZF-${\mu}$SR spectrum is characterised by a weak depolarization of the muon spin ensemble and show no evidence of long-range ordered magnetism in ScV$_{6}$Sn$_{6}$. However, it shows that the muon spin relaxation has a clearly observable temperature dependence. Figure 3$\bf{c}$ displays the ${\mu}$SR spectra collected at 5 K in zero-field and at various external magnetic fields applied longitudinal to the muon spin polarization, $B_{{\rm LF}}$ ${\simeq}$ 1-25 mT. Since the full polarization can be recovered  by the application of a small external longitudinal magnetic field, $B_{{\rm LF}}$~=~5~mT, the relaxation is, therefore, due to spontaneous fields which are static on the microsecond timescale \cite{Yaouanc}. Similar to AV$_{3}$Sb$_{5}$, the zero-field ${\mu}$SR spectra for ScV$_{6}$Sn$_{6}$ were fitted using the gaussian Kubo-Toyabe depolarization function \cite{Toyabe}, multiplied by an additional exponential exp(-$\Gamma t$) term

\begin{figure*}[t!]
\centering
\includegraphics[width=0.75\linewidth]{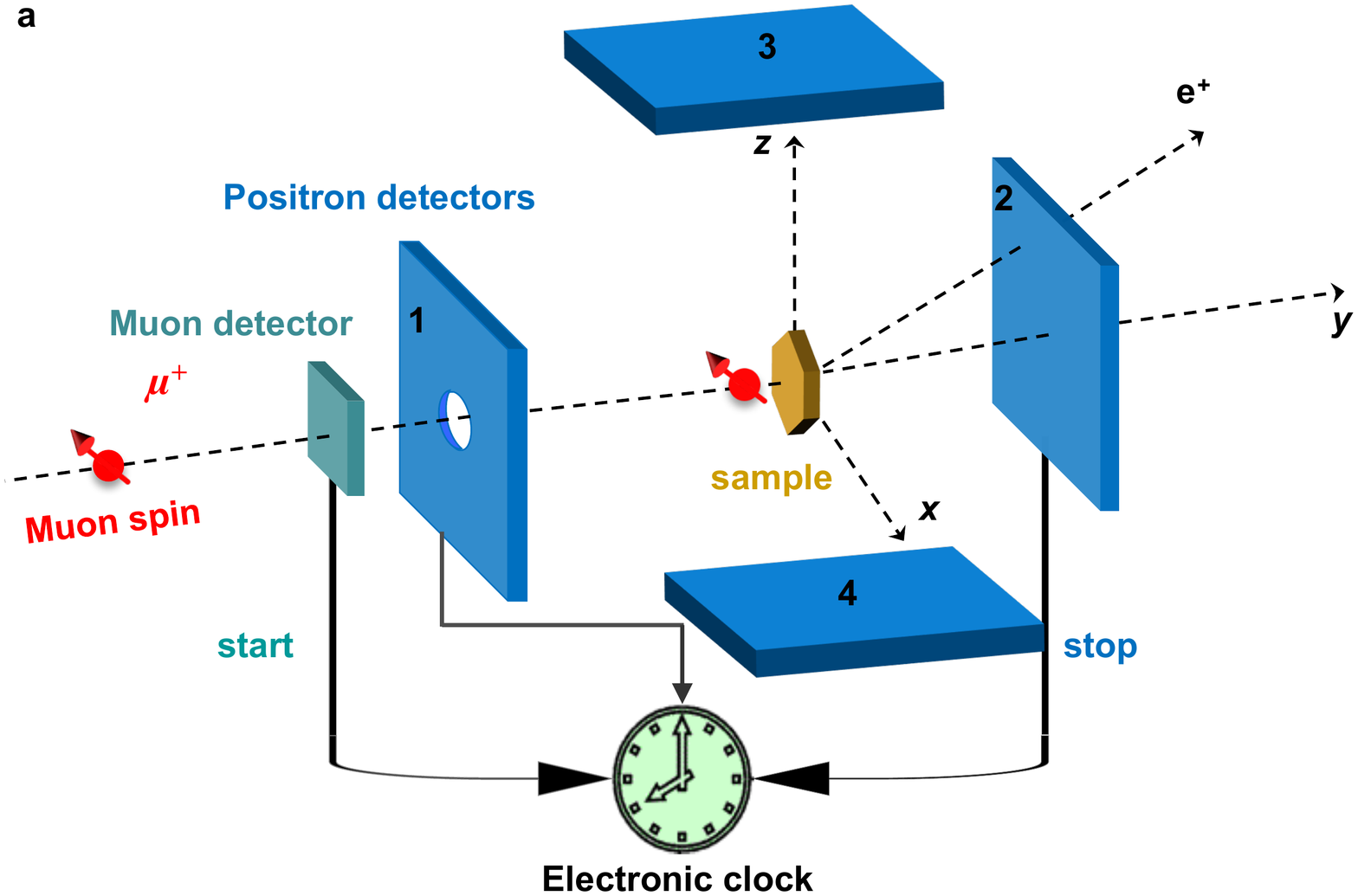}
\includegraphics[width=0.75\linewidth]{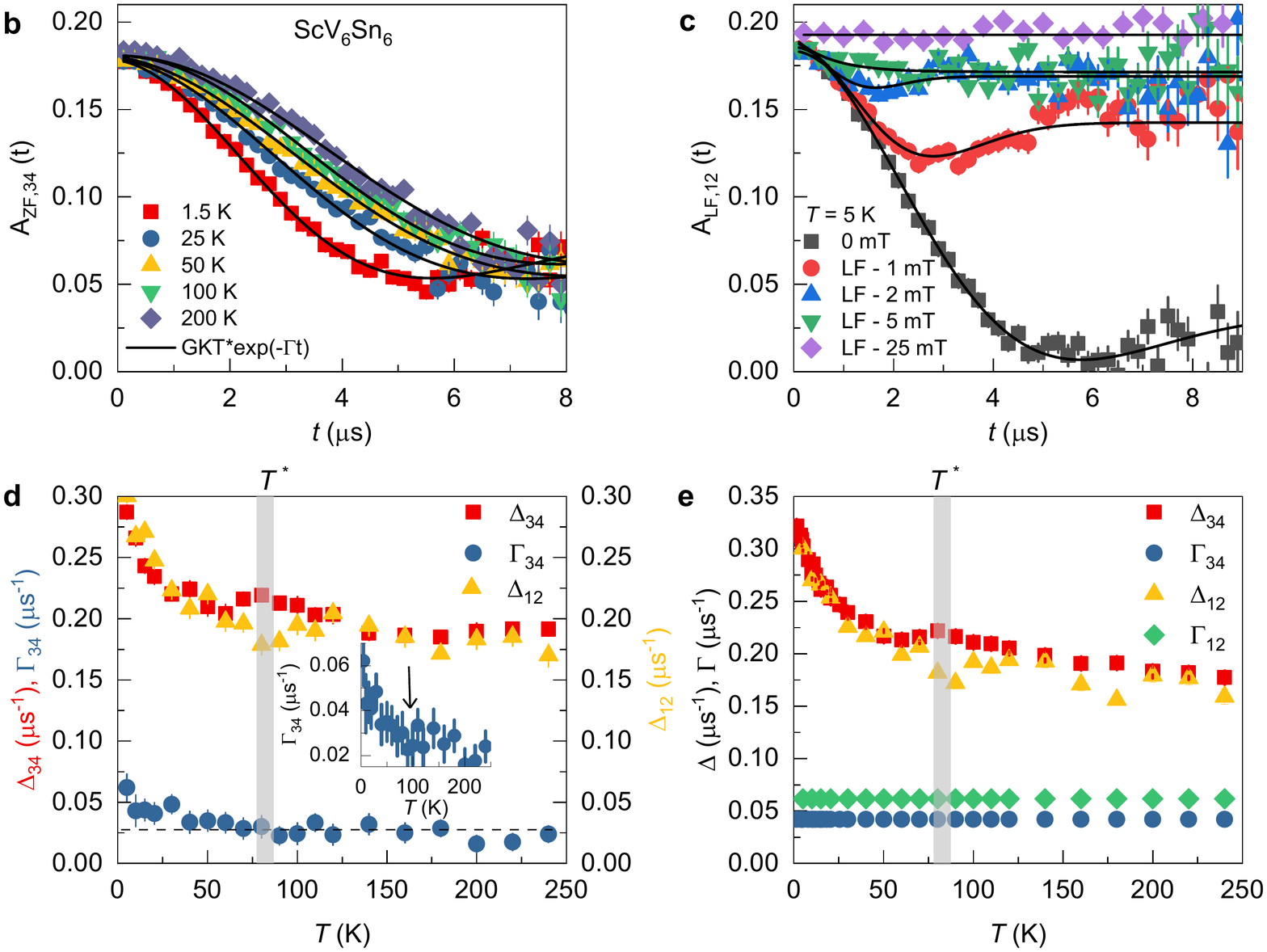}
\vspace{-0.3cm}
\caption{\textbf{Magnetic response of the charge order in ScV$_{6}$Sn$_{6}$.} 
$\bf{a,}$ Overview of the experimental setup. Spin polarized muons with spin $S_{\mu}$, forming $45^{\circ}$ with respect to the $c$-axis of the crystal, are implanted in the sample, which is surrounded by four positron detectors: Forward (1), Backward (2), Up (3), and Down (4).
A clock is started at the time the muon surpasses the muon detector (M) and is stopped as soon as the decay positron is detected in one of the detectors. $\bf{b,}$ The ZF ${\mu}$SR time spectra for ScV$_{6}$Sn$_{6}$, obtained at different temperatures. The solid black curves in panel $\bf{b}$ represent fits to the recorded time spectra, using the Eq. 1. Error bars are the standard error of the mean (s.e.m.) in about 10$^{6}$ events. $\bf{c,}$ The ${\mu}$SR time spectra for ScV$_{6}$Sn$_{6}$, obtained at $T$ = 5 K in zero-field and at various external magnetic field applied longitudinal to the muon spin polarization. The solid black curves in panel c represent fits to the recorded time spectra, using the Eq. 1. Error bars are the standard error of the mean (s.e.m.) in about 10$^{6}$ events. $\bf{d-e,}$ The temperature dependences of the relaxation rates ${\Delta}$ and ${\Gamma}$  from two sets of detectors, obtained in a wide temperature range across the charge ordering temperature $T^{*}$ ${\simeq}$ 80 K. In panel $\bf{e}$ the rate ${\Gamma}$ is kept constant as a function of temperature. The error bars represent the standard deviation of the fit parameters.}
\label{fig1}
\end{figure*}


\begin{equation}
\begin{aligned}
P_{ZF}^{GKT}(t) =  \left(\frac{1}{3} + \frac{2}{3}(1 - \Delta^2t^2 ) \exp\Big[-\frac{\Delta^2t^2}{2}\Big]\right) \exp(-\Gamma t) \\ 
\end{aligned}
\end{equation}  
where ${\Delta}$/${\gamma_{\mu}}$ is the width of the local field distribution primarily due to the nuclear moments. However, this Gaussian component may also include the field distribution at the muon site created by a dense network of weak electronic moments. ${\gamma_{\mu}}$/2${\pi}$ = 135.5~MHz/T is the muon gyromagnetic ratio. The exponential term ${\Gamma}$ may e.g. be due to the presence of electric field gradients, causing deviations from the GKT-like spectrum or dilute electronic moments. ${\Delta}_{34}$ shows a non-monotonous temperature dependence; namely, a peak coinciding with the onset of the charge order, which decreases to a broad minimum before increasing again towards lower temperatures. ${\Delta}_{12}$ instead shows a weak minimum at $T^{*}$ with the significant increase at lower temperatures. The onset of charge order might alter the electric field gradient experienced by the nuclei, due to the fact that the quantization axis for the nuclear moments depends on the electric field gradients (EFG), and correspondingly the magnetic dipolar coupling of the muon to the nuclei \cite{Sonier2012}. This can induce a change in the nuclear dipole contribution to the zero-field ${\mu}$SR signal and may explain the small maximum or minimum in ${\Delta}_{34}$ and ${\Delta}_{12}$, respectively, at the onset of $T^{*}$. However, the significant increase of both ${\Delta}_{34}$ and ${\Delta}_{12}$ at lower temperatures is difficult to explain with the change of the EFG and suggests a considerable contribution of electronic origin (dense moments) to the muon spin relaxation in the charge ordered state. There is also a noteworthy increase in the relaxation rate ${\Gamma}$$_{34}$ upon lowering the temperature below the charge ordering temperature $T^{*}$, which is better visible in the inset of Fig.~3$\bf{d}$. Moreover, our high field ${\mu}$SR results presented below definitively prove that there is indeed a strong contribution of electronic origin to the muon spin relaxation below the charge ordering temperature (see below). In sum, the ZF-${\mu}$SR results indicate that there is an enhanced width of internal fields sensed by the muon ensemble below $T^{*}$ ${\simeq}$ 80 K. The increase of ${\Delta}$ and ${\Gamma}$ below $T^{*}$ as well as the dip-like feature around $T^{*}$ are reminiscent of the behavior observed in other kagome systems AV$_{3}$Sb$_{5}$ (A = K, Rb, Cs). The increase in ${\Delta}_{34}$ and ${\Delta}_{12}$ below $T^{*}$ in  ScV$_{6}$Sn$_{6}$  is estimated to be ${\simeq}$ 0.1~${\mu}$$s^{-1}$ and ${\simeq}$ 0.12~${\mu}$$s^{-1}$, respectively, which can be interpreted as a characteristic field strength ${\Delta}$$_{34,12}$/${\gamma_{\mu}}$ ${\simeq}$ 1.2-1.5~G. We note that this value is by factor of ${\sim}$ 4-5 higher than the one reported previously for AV$_{3}$Sb$_{5}$ (A = K, Rb, Cs). A similar increase of internal magnetic field strength is reported in several time-reversal symmetry-breaking superconductors \cite{LukeTRS} and in some multigap TRS breaking superconductors (e.g. La$_{7}$Ni$_{3}$ \cite{Singh}) across $T_{{\rm c}}$. 

\begin{figure*}[t!]
\includegraphics[width=0.9\linewidth]{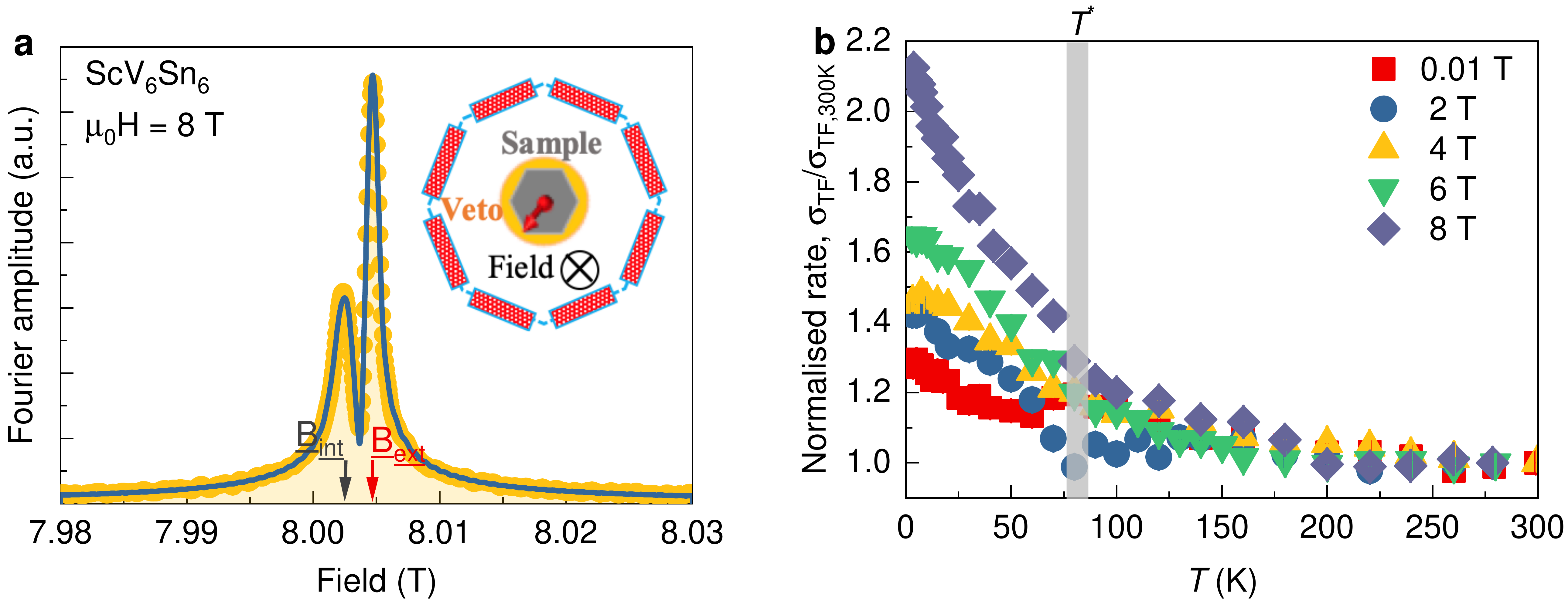}
\vspace{-0.0cm}
\caption{\textbf{Enhanced magnetic response of the charge order with applying external magnetic fields.} 
$\bf{a,}$ Fourier transform of the ${\mu}$SR asymmetry spectra for a mosaic of single crystals of ScV$_{6}$Sn$_{6}$ at 3 K in the presence of an applied field of ${\mu}_{0} H = 8 $T. The black solid line is a two-component signal fit. The peaks marked by the arrows denote the external and internal fields, determined as the mean values of the field distribution from the silver sample holder and from the sample, respectively. Inset shows the schematic high-field ${\mu}$SR experimental setup. The sample was surrounded by 2 ${\times}$ 8 positron detectors, arranged in rings. The specimen was mounted in a He gas-flow cryostat with the largest face perpendicular to the muon beam direction, along which the external field was applied. Behind the sample lies a veto counter (in orange) which rejects the muons that do not hit the sample. $\bf{b,}$ Temperature dependence of the high transverse field muon spin relaxation rate $\sigma_{\rm TF}$ for the single crystal of ScV$_{6}$Sn$_{6}$, normalized to the value at 300 K, measured under different $c$-axis magnetic fields. The vertical grey lines mark the charge ordering temperature, determined from specific heat measurements. The error bars represent the standard deviation of the fit parameters.}
\label{fig3}
\end{figure*}

In order to substantiate the zero-field ${\mu}$SR results, presented above, we carried out systematic high field ${\mu}$SR \cite{Sedlak} experiments. Under high magnetic field, the direction of the applied field defines the quantization axis for the nuclear moments, so that the effect of the charge order on the electric field gradient at the nuclear sites is irrelevant. For the high-field experiments, a mosaic of several crystals was used. The individual crystals were glued to a 10 mm circular silver sample holder and the entire ensemble was held together by small droplets of GE varnish. Figure 4$\bf{a}$ shows the probability field distribution, measured at 3 K in a magnetic field of 8 T applied along the crystallographic $c$-axis. In the whole investigated temperature range, two-component signals were observed: a signal with fast relaxation $\sigma_{\rm TF}$ ${\simeq}$ 0.428(3) ${\mu}s^{-1}$  (broad signal on the left side of the Fourier spectrum) and another one with a slow relaxation 0.05 ${\mu}s^{-1}$ (narrow signal). The narrow signal arises mostly from the muons stopping in the silver sample holder and its position is a precise measure of the value of the applied magnetic field. The width and the position of the narrow signal is found to be temperature independent, as expected, and thus were kept constant in the analysis. The relative fraction of the muons stopping in the sample was fixed to the value obtained at the base-$T$ and kept temperature independent. The signal with the fast relaxation, which is shifted towards a lower field from the applied one, arises from the muons stopping in the sample and it takes a major fraction (${\sim}$ 50 ${\%}$) of the ${\mu}$SR signal. This points to the fact that the sample response arises from the bulk of the sample. A non-monotonous behaviour of the relaxation rate is clearly seen in the ${\mu}$SR data, measured in magnetic field of 0.01~T, applied parallel to the $c$-axis, as shown in Fig.~4$\bf{b}$. This looks similar to the temperature dependence of the zero-field rate ${\Delta}_{34}$. At higher fields such as 2~T, 4~T, 6~T, and 8~T, the rate shows a clear and stronger increase towards low temperatures within the charge ordered state. As the nuclear contribution to the relaxation cannot be enhanced by an external field, this indicates that the low-temperature relaxation rate in magnetic fields is dominated by the electronic contribution. Remarkably, we find that the absolute increase of the relaxation rate between  the onset of charge order $T^{*}$ and the base-$T$ in 8~T is  ${\Delta}$$\sigma_{\rm TF}$ ${\simeq}$ 0.23 ${\mu}$$s^{-1}$ which is a factor of two higher than the one 0.12 ${\mu}$$s^{-1}$ observed in zero-field. The transverse-field relaxation is normally smaller by a factor of ${\sim}$ 2 than expected from the zero-field measurements \cite{Yaouanc}. This is because in transverse-field measurements only the width in one direction is relevant while in ZF the width in two directions is relevant. This means that the relaxation for ScV$_{6}$Sn$_{6}$ in high transverse-field is actually a factor of four higher than in ZF. This indicates a strong field-induced enhancement of the electronic response. We also note that a weak but non-negligible field effect on the relaxation rate is observed above $T^{*}$ within a 20-30 K temperature range, which may point towards the charge density wave fluctuations preceding the phase transition. 


Based on the unique combination of ZF-${\mu}$SR and high-field ${\mu}$SR experiments, we observed a magnetic response (enhanced internal field width) below $T^{*}$ ${\simeq}$ 80 K, providing direct evidence for the time-reversal symmetry-breaking fields in the kagome lattice of ScV$_{6}$Sn$_{6}$. Since at least 50 ${\%}$ of the sample volume experiences an increase in the relaxation rate, this indicates the bulk nature of the transition below $T^{*}$. The presence of an electronic response and its field induced enhancement in ScV$_{6}$Sn$_{6}$ is similar to our previous observations in kagome lattice superconductors  $A$V$_{3}$Sb$_{5}$ ($A$ = K, Rb, Cs)\cite{GuguchiaMielke,GuguchiaRVS,KhasanovCVS}. 

We specify some similarities and differences between ScV$_{6}$Sn$_{6}$ and $A$V$_{3}$Sb$_{5}$: (1) In ScV$_{6}$Sn$_{6}$, we observed the enhancement of the rate ${\Gamma}$$_{34}$ by 0.03 ${\mu}$$s^{-1}$ below $T^{*}$, which is similar to KV$_{3}$Sb$_{5}$. 
On the other hand, the Gaussian rate ${\Delta}$$_{34}$ and ${\Delta}$$_{12}$ show significant increase, i.e. by 0.12 ${\mu}$$s^{-1}$, in ScV$_{6}$Sn$_{6}$, while it shows only a very weak temperature dependence at low temperatures in $A$V$_{3}$Sb$_{5}$.
These results show that the increase of relaxation in ScV$_{6}$Sn$_{6}$ is due predominantly to Gaussian rate.   
This indicates that internal fields arise from a network of dense weak electronic moments in ScV$_{6}$Sn$_{6}$, while the electronic moments seem to be more dilute or more dynamic in $A$V$_{3}$Sb$_{5}$. (2) Among the three compounds $A$V$_{3}$Sb$_{5}$ ($A$ = K, Rb, Cs) the largest increase of the zero-field relaxation rate below $T^{*}$ was observed for RbV$_{3}$Sb$_{5}$. In ScV$_{6}$Sn$_{6}$, the increase of the zero-field relaxation is 0.12 ${\mu}$$s^{-1}$, which is a factor of two higher than the one 0.06 ${\mu}$$s^{-1}$ observed in RbV$_{3}$Sb$_{5}$. (3) The largest field effect was observed in KV$_{3}$Sb$_{5}$. In ScV$_{6}$Sn$_{6}$, magnetic field also leads to the enhancement of the rate and the increase of the rate reaches a value of 0.23 ${\mu}$$s^{-1}$ at 8 T. This is a factor of 1.6 higher than the one (0.15 ${\mu}$$s^{-1}$) observed in KV$_{3}$Sb$_{5}$. The $A$V$_{3}$Sb$_{5}$ systems are close to the condition of a kagome lattice with van Hove filling and with extended Coulomb interactions \cite{MDenner,GuguchiaMielke}. TRS breaking charge order in $A$V$_{3}$Sb$_{5}$ was interpreted in terms of orbital current order \cite{MDenner,MHChristensen,Balents,Nandkishore,DSong,MortenRafael}, which may fundamentally affect the superconducting state.  
Muons couple to the closed current orbits below $T^{*}$, leading to an enhanced internal field width sensed by the muon ensemble concurrent with the charge order \cite{GuguchiaMielke,GuguchiaRVS}. So far, no theoretical proposal for the orbital current order was reported for ScV$_{6}$Sn$_{6}$. In addition, recent STM and ARPES results \cite{ManuelT,ChengS} reveal that the nature of the CDW order in ScV$_{6}$Sn$_{6}$ is different from $A$V$_{3}$Sb$_{5}$. With the lack of theoretical understanding of ScV$_{6}$Sn$_{6}$, we cannot conclude on the origin of the TRS breaking field in this system. But, our results provide key evidence that the magnetic and charge channels of ScV$_{6}$Sn$_{6}$ are strongly intertwined and will inspire future experiments, particularly neutron scattering with polarization analysis, to potentially understand the precise origin of the observed magnetism.

The exploration of unconventional electronic phases that result from strong electronic correlations is a frontier in condensed matter physics. Kagome lattice systems appear to be an ideal setting in which strongly correlated topological electronic states may emerge.
Most recently, the kagome lattice of $A$V$_{3}$Sb$_{5}$ has shown to host unconventional chiral charge order, which is analogous to the long-sought-after quantum order in the Haldane model \cite{Haldane} for honeycomb lattice or Varma model \cite{Varma} for cuprate high-temperature superconductors. Here we employ muon spin relaxation to probe the magnetic response of kagome charge order on a microscopic level in newly discovered ScV$_{6}$Sn$_{6}$ with a vanadium kagome lattice. We found an enhancement of the internal field width sensed by the muon ensemble, which takes place within the charge ordered state. The muon spin relaxation rate below the charge ordering temperature is substantially enhanced by applying an external magnetic field. Our work points to a time-reversal symmetry-breaking charge order in ScV$_{6}$Sn$_{6}$ and extends the classification of materials with unconventional charge order beyond the series of compounds $A$V$_{3}$Sb$_{5}$ ($A$ = K, Rb, Cs).

\section{METHODS}

\textbf{Sample preparation}: Single crystals of ScV$_{6}$Sn$_{6}$ were grown by the molten Sn flux method. The starting materials were Sc (3N, Alfa Aesar), V (2N8, Alfa Aesar), and Sn (99.9999 ${\%}$, Alfa Aesar). In a helium-filled glovebox, Sc (0.218 g, 0.485 mmol), V (1.481 g, 2.908 mmol), and Sn (33.390 g, 28.127 mmol) were placed into a 5 ml alumina Canfield crucible \cite{Canfield}. The crucible set was placed into a quartz u-tube, evacuated, backfilled with c.a. 100 mbar of Ar, and sealed as an ampoule.  The specimen was heated up to 1150 °C, with a rate of 400 °C/h, and annealed at that temperature for 12 h.  Subsequently, the sample was cooled down to 780 °C with a rate of 100 °C/h.  After the growth step, the excess Sn flux was separated from the single crystals by centrifugation.\\ 

\textbf{Crystallography}: A single crystal of ScV$_{6}$Sn$_{6}$, obtained from Sn flux, was mounted on the $MiTeGen~MicroMounts$ loop and used for a X-ray structure determination.  Measurements were performed at RT on a $STOE~STADIVARI$ diffractometer equipped with a $Dectris~ EIGER~1M~2R~CdTe$ detector and with an $Anton~Paar~Primux$ 50 Ag/Mo dual-source using Ag $K_{\rm \alpha}$ radiation (${\lambda}$ = 0.56083 Å) from a micro-focus X-ray source and coupled with an $Oxford~Instruments~Cryostream$ 800 jet.  The unit cell constants and an orientation matrix for data collection were obtained from a least-squares refinement of the setting angles of 16647 reflections in the range 6.7° ${\textless}$ 2${\theta}$ ${\textless}$ 66.6°.  A total of 3448 frames were collected using ${\omega}$ scans, 5 seconds exposure time and a rotation angle of 0.5° per frame, and a crystal-to-detector distance of 60.0 mm.\\ 

Data reduction was performed with $X$-$Area$ package [$X$-$Area$ package, Version 2.1, STOE and Cie GmbH, Darmstadt, Germany, 2022].  The intensities were corrected for Lorentz and polarization effects, and an empirical absorption correction using spherical harmonics was applied [X-Area package, Version 2.1, STOE and Cie GmbH, Darmstadt, Germany, 2022].  The structure was solved using $ShelXT$ \cite{SheldrickT} and $Olex$$s^{2}$ program \cite{Dolomanov}.  The model was refined with $ShelXL$ package \cite{SheldrickL} and $Olex$$s^{2}$ software.\\  

\textbf{Heat Capacity}: Heat Capacity ($C_{\rm p}$) measurements were performed in a Physical Properties Measurement System (PPMS, 14T, Quantum Design) at zero magnetic field using the relaxation method between 1.85 and 200 K. 7.75 mg of ScV$_{6}$Sn$_{6}$ crystal was fixed with the Apiezon-N grease to the sapphire holder of the calorimeter. The samples were then cooled down to 1.85 K, and the heat capacity measured by heating.  The signal from the Apiezon-N grease and the calorimeter, previously measured under the same conditions, was subtracted from the data to obtain the sample`s $C_{\rm p}$.\\  

\textbf{Magnetotransport}: Magnetoresistance of the single-crystal ScV$_{6}$Sn$_{6}$ was measured in physical property measurement system (PPMS-9, Quantum Design) using the typical four-probe technique. Four Pt-wires (0.0254 mm diameter) were attached by silver epoxy on the single crystal, polished into bar-shaped specimen. Electrical current of 1 mA and magnetic field was applied along the crystallographic $a$- and $c$-axis, respectively, that were confirmed by Laue measurement.\\ 

\textbf{${\mu}$SR experiment}: In a ${\mu}$SR experiment nearly 100 ${\%}$ spin-polarized muons ${\mu}$$^{+}$
are implanted into the sample one at a time. The positively charged ${\mu}$$^{+}$ thermalize at interstitial lattice sites, where they
act as magnetic microprobes. In a magnetic material the muon spin precesses in the local field $B_{\rm \mu}$ at the
muon site with the Larmor frequency ${\nu}_{\rm \mu}$ = $\gamma_{\rm \mu}$/(2${\pi})$$B_{\rm \mu}$ (muon
gyromagnetic ratio $\gamma_{\rm \mu}$/(2${\pi}$) = 135.5 MHz T$^{-1}$).\\

Zero field (ZF) and transverse field (TF) $\mu$SR experiments on the single crystalline sample of ScV$_{6}$Sn$_{6}$ were performed on the GPS instrument and high-field HAL-9500 instrument, equipped with BlueFors vacuum-loaded cryogen-free dilution refrigerator (DR), at the Swiss Muon Source (S$\mu$S) at the Paul Scherrer Institut, in Villigen, Switzerland. A mosaic of several crystals stacked on top of each other was used for these measurements. The magnetic field was applied along the crystallographic $c$-axis. A schematic overview of the 
experimental setup for zero-field is shown in Figure 4a. The crystal was mounted such that the $c$-axis of it is parallel to the muon beam. Using the ''spin rotator'' at the ${\pi}$M3 beamline, muon spin was rotated (from its natural orientation, which is antiparallel to the momentum of the muon) by $44.5(3)^{\circ}$ degrees with respect to the $c$-axis of the crystal. So, the sample orientation is fixed but the muon spin was rotated. The rotation angel can be precisely estimated to be $44.5(3)^{\circ}$ by measurements in weak magnetic field, applied transverse to the muon spin polarization. Zero field and high transverse field $\mu$SR data analysis on single crystals of ScV$_{6}$Sn$_{6}$ were performed using both the so-called asymmetry and single-histogram modes \cite{Suter,Yaouanc}. The experimental data were analyzed using the MUSRFIT package \cite{Suter}.\\

\textbf{Competing interests:} All authors declare that they have no competing interests.\\

\textbf{Data Availability}: All relevant data are available from the authors. Alternatively, the data can be accessed through the data base at the following link http://musruser.psi.ch/cgi-bin/SearchDB.cgi.\\

\section{Acknowledgments}~
The ${\mu}$SR experiments were carried out at the Swiss Muon Source (S${\mu}$S) Paul Scherrer Insitute, Villigen, Switzerland.
Z.G. acknowledges support from the Swiss National Science Foundation (SNSF) through SNSF Starting Grant (No. TMSGI2${\_}$211750).
Z.G. acknowledges the useful discussions with Robert Scheuermann. Y.H. and  M.S. acknowledges support from the Swiss National Science Foundation under Grant. No. 200021${\_}$188413.

\end{document}